\documentclass[submission,copyright,creativecommons]{eptcs}
\providecommand{\event}{AiSoS 2013} 
\usepackage{breakurl}             
\usepackage{xcolor,colortbl}
\usepackage[normalem]{ulem}

\title{Verification and Validation Issues in Systems of Systems}
\author{Dr. Eric Honour
\institute{Honourcode, Inc.\\
1169 McCoury Ln\\
Spring Hill, TN 37174 USA}
\email{ehonour@hcode.com}
}
\def\titlerunning{Verification and Validation Issues in Systems of Systems}
\def\authorrunning{Dr. Eric Honour}
\begin{document}
\maketitle

\begin{abstract}
The cutting edge in systems development today is in the area of “systems of systems,” (SoS) large networks of inter-related systems that are developed and managed separately, but that also perform collective activities.  Such large systems typically involve constituent systems operating with different life cycles, often with uncoordinated evolution.  The result is an ever-changing SoS in which adaptation and evolution replace the older engineering paradigm of “development.”  This short paper presents key thoughts about verification and validation in this environment.  Classic verification and validation methods rely on having (a) a basis of proof, in requirements and in operational scenarios, and (b) a known system configuration to be proven.  However, with constant SoS evolution, management of both requirements and system configurations are problematic.  Often, it is impossible to maintain a valid set of requirements for the SoS due to the ongoing changes in the constituent systems.  Frequently, it is even difficult to maintain a vision of the SoS operational use as users find new ways to adapt the SoS.   These features of the SoS result in significant challenges for system proof.  In addition to discussing the issues, the paper also indicates some of the solutions that are currently used to prove the SoS.
\end{abstract}

\section{Systems of Systems}

The concept of a system of systems (SoS) has developed in the last decade and a half through recognition that many systems are now interconnected to achieve functionality and characteristics that cannot be achieved by the individual constituent systems.  Yet the SoS concept is still poorly defined.  Some definitions include:

\begin{quote}``An SoS is a system comprised of elements that are systems.'' (Widespread understanding; assumed in Manthorpe 1996 \cite{manthorpe1996})\end{quote}

\begin{quote}``Systems of Systems are large-scale concurrent and distributed systems that are comprised of complex systems'' (Jamshidi 2005 \cite{jamshidi2005system})\end{quote}

\begin{quote}``An SoS is defined as a set or arrangement of systems that results when independent and useful systems are integrated into a larger system that delivers unique capabilities'' (US DoD 2008, SoSE Guide \cite{DoDSoSguide2008})\end{quote}

These definitions appear to apply to common SoS examples such as military systems, a modern airport, or manufacturing supply chain management.  In these examples, many constituent systems interconnect to obtain high-level functionality that exists above and beyond the individual systems.  In these examples, the constituent systems are typically developed and upgraded in different time frames by different system owners, adding to the difficulty of the SoS.  But the weakness in these definitions is that they might also apply to other systems, such as a microprocessor, that are clearly not included in the usual SoS considerations.  These definitions are useful, but represent muddy thinking that does not distinguish between a ``system'' and an ``SoS''.

The best distinction available was proposed by \cite{maier1998architecting}, in which a system is considered an SoS if it exhibits significant amounts of:
\begin{itemize}
\item{\bf Emergent Behavior.}  The SoS performs functions that are not achievable by the independent constituent systems; stakeholders want assurance of this emergent behavior even in the face of the challenges created by the following characteristics.
\item{\bf Geographic Distribution.} Spreading the constituent systems across a geographic extent forces the elements to exchange information in a remote way, resulting in difficult technical communications issues.
\item{\bf Evolutionary Development.}  Functions and purposes are added, removed, and modified within the system in an ongoing way.  As the constituent systems change, there are constant revisions and difficult integration issues.
\item{\bf Operational Independence.}  The constituent systems have purpose, even if detached from the SoS.  The purposes often conflict with each other and also conflict with the purposes of the SoS, resulting in conflicts among the constituent system stakeholders.
\item{\bf Managerial Independence.}  The constituent systems are developed and managed for their own purposes.  Each system has an independent owner and independent stakeholders, who may or may not overlap with the SoS stakeholders.  This independence further exacerbates the conflicts.
\end{itemize}

The result of these characteristics is a need for new paradigms in systems engineering (SE).  Many traditional SE concepts are counter-productive to the SoS, resulting in methodology that is more evolutionary than developmental in nature, with constant evaluation that results in suggested or influenced changes in the constituent systems.  In many cases, there is actually no central manager for the SoS, so all changes are handled in a collaborative or competitive way among the constituent system owners.

\section{Traditional Approach to Verification and Validation}
The long history of systems development has led to a well-understood practice of the proof of completion.  This practice is encompassed in two separate approaches, both of which are part of normal system development:
\begin{itemize}
\item{\bf Verification} is checking whether an item or artifact has been designed/built to conform with requirements.  Multiple methods may be used for the checking, such as inspection, analysis, demonstration, or test.  This can be applied to the final system, or it can be applied to design artifacts produced during the system development.  An underlying necessity is to have a known list of requirements stated in such a representation as to be firmly verifiable.
\item{\bf Validation} is checking whether an item or artifact has been designed/built to fulfill its intended purpose.  Again, multiple methods may be applied to the system or to design artifacts.  However, the standard against which validation is performed is the mission need as perceived by the stakeholders.  This standard is often subjective in nature, truly known only by the stakeholders and often only known after the fact.
\end{itemize}

Verification is the basis of the normal ``Vee'' model of systems engineering, in which requirements are progressively analyzed and allocated downward into lower levels of the system hierarchy, followed by verification actions against those requirements while building up the system.  If the Vee model is rigorously followed with firm, verifiable requirements, then the resulting system can be theoretically shown to conform completely with its requirements.

Validation is also necessary, however, because the true need for the system cannot usually be expressed in firm, verifiable requirements.  The translation from the stakeholder desires into system requirements is performed through operations analysis and is never perfect.  Therefore, multiple levels of requirements validation are necessary to help the requirements conform to the stakeholder need.  At the completion of development, a system validation is necessary to ensure that the final system actually performs as the stakeholders desire.  And during the development, mission validation is also necessary to encompass any changes that occur in the stakeholders’ perceptions.

These methods work reasonably well in normal system development.  The characteristics of an SoS, however, create issues that act to the detriment of these methods.

\section{Systems of Systems Verification}
The verification approach to system proof presupposes a known standard against which to verify.  That standard must be precise enough to act as a verifiable standard.  For normal systems it is usually in the form of requirements that are agreed or mandated by an authority for the system.  In traditional systems engineering, requirements serve six major purposes:
\begin{itemize}
\item{\bf Agreement.}  Requirements define the technical agreements between an acquirer and a supplier, between stakeholders and developers, and between developers and testers.
\item{\bf Contract bounds.}  Requirements serve as the technical definition of a contract for system development.
\item{\bf Analysis.}  Requirements provide the tool for analysis of the coherence, completeness, and correctness of the technical definition.  By analyzing the requirements against each other and against the stakeholder needs, the technical definition can be made as assured as possible.
\item{\bf In-process verification.}  Requirements provide the standard against which to verify the in-process artifacts of design, as a way to check the design during the development process.
\item{\bf Allocation.}  Requirements are allocated downward to lower levels of the system definition, with bidirectional traceability to ensure that lower levels perform all the work and only the work necessary for the system.
\item{\bf System verification.}  Requirements provide the standard against which to verify the completed system.
\end{itemize}

These are all useful purposes, which is why requirements have served well for many decades of system development.  However, in the case of the SoS, there are issues with every one of these purposes that derive from the nature of the SoS itself, as shown in the table below.  Even in the case of normal systems, recent advances in Agile development have been questioning the utility of requirements for evolutionary progress.

\colorlet{brightgray}{gray!20}
{\renewcommand{\arraystretch}{1.2}%
\begin{tabular}{p{4.6cm}||p{4.6cm}||p{4.6cm}}

\rowcolor{black}
\color{white}\bf Task &
\color{white}\bf System &
\color{white}\bf SoS \\

\rowcolor{lightgray}
Define agreement\newline 
~-~Acquirer/supplier\newline 
~-~Stakeholders/\newline
~-~developers/testers &
All apply; roles are known; agreement can happen &
Conflicting goals; evolutionary growth; agreement is very difficult \\

\rowcolor{brightgray}
Define the technical bounds on a contract &
Used for contracting &
Typically NO SoS contract \\

\rowcolor{lightgray}
Analysis for coherence, \newline complete, correct &
Analysis as part of contract work &
Conflicting ideas of\newline ``complete,'' ``correct'' \\

\rowcolor{brightgray}
Checking the design during development &
Req’s used at all reviews &
Development happens at CS level \\

\rowcolor{lightgray}
Allocation, trace into lower levels &
Part of system design work &
Only possible if SoS level has been agreed \\

\rowcolor{brightgray}
Checking each item at \newline completion &
System verification;\newline component verification &
Only possible if req's have been agreed \\
\end{tabular}}
~\newline

One solution for SoS {\em verification} is to use the constituent system {\em validation} events, commonly called ``operational test \& evaluation (OT\&E)'' to verify the SoS functionality against the perceived SoS requirements.  This provides the advantage of exercising the SoS in a ``real world'' setting using experienced human operators, but suffers the disadvantages of lack of firm control of the environment, lack of repeatability, and lack of ``ground truth'' for the proof.
The concept of verification is exceedingly difficult for the SoS, primarily because of the need for firm requirements against which to verify.  This need is counter-indicated by the conflicting, evolutionary goals of the normal SoS.  In one example of this difficulty, the US Army’s Future Combat Systems attempted to apply firm flow-down of requirements into the largest procurement the US Department of Defense has run.  The environment and needs evolved more rapidly than the development team could manage.  As a result, in six years of work the projected initial deployment date moved {\em seven} years outward.  The DoD cancelled the effort.

\section{Systems of Systems Validation}
Validation in the SoS presents its own set of difficulties, but is largely more tractable than verification.  There are multiple levels of test and evaluation in any SoS, that happen at asynchronous times:
\begin{itemize}
\item{\bf Interface certification testing} frequently provides a means to qualify the constituent systems for operation within the SoS.
\item{\bf Constituent system developmental testing} performs verification events on the constituent systems, controlled by the individual system owners and developers.
\item{\bf Constituent system operational testing} performs validation on the constituent systems, often within the actual SoS environment.
\item{\bf SoS simulations} provide a method to execute some of the SoS operations in a controlled environment, although the accuracy of the simulations is always suspect.
\item{\bf SoS testing} typically occurs in situ during operation of the SoS in its real environment.  In some cases, it may be possible to perform testing during training events or preliminary operations.  This is often the only method to test and/or discover SoS emergent behaviors.
\end{itemize}

The test responsibilities for these various events resides with different organizations, carrying through the same SoS issues with managerial independence that create conflicting goals and conflicting desires.  Each constituent system has its own test responsibilities, with the test focus on the functions and performance of that system.  They may be motivated to be part of the SoS, but proof of the SoS is beyond the scope of the constituent system test owners.  There may or may not be a test owner at the SoS level who has focus on the SoS testing.

Test and evaluation support to the SoS, however, is ongoing and nearly continuous.  Each change to a constituent system results in test events that affect the SoS.  Each test event can provide key information to guide the SoS engineering.

Because of the dynamic, evolutionary nature of the SoS, continuous validation of the SoS functionality and performance is a viable approach to proof.  As real-life events occur in operation, instrumentation of the SoS response provides operating testing results that can feed back into the SoS engineering cycle.  In some cases, events may even be created in a controlled way to test specific functionality.

However, many SoS scenarios cannot be executed in real life due to various dangers to humans or systems.  In other cases, test scenarios may prevent the SoS from operating in essential ways.  In such cases, the best approach seems to be the use of SoS simulations.  Such simulations require care.  In many cases, creating the simulations may be an extensive engineering effort.  For any type of accuracy, the simulations must reflect the stakeholder performance issues, conflicting goals, patterns, and emergent behavior (both designed and unknown).

The validation approach to SoS proof recognizes the essentially subjective nature of the conflicting stakeholder goals.  An evolutionary development approach that embraces this nature can work well.  An example is the Global Earth Observation SoS (GEOSS).  Envisioned in 2002 as a multi-national effort, GEOSS has been working to connect and pool information from many disparate Earth observation systems.  Guided by an executive consortium of 12 members, the technical approach has been to create interoperability while recognizing the charters of the individual systems.  Work continues over a decade later, with significant progress.

\section{Conclusions}
The nature of the SoS creates significant difficulties for system-level proof.  In traditional systems engineering, proof is created through verification and validation approaches.

Verification approaches, when applied to the SoS, have problems creating agreement on the standard against which to verify.  Conflicting goals, lack of an SoS authority, and evolutionary growth all contribute to an inability to use requirements in the same way as in normal systems.  There are also problems with formal verification methods due to the informal nature of the SoS requirements, the lack of ability to control SoS events, and the evolutionary growth that makes test planning difficult.  System validation events (operational test \& evaluation) provide some ability to approach a form of SoS verification.

Validation approaches applied to the SoS appear to have better viability than verification approaches.  Validation recognizes the essentially subjective nature of the conflicting, evolutionary goals, allowing the SoS configuration and proof to adapt over time as needed.  Within the context of the ongoing SoS, in situ evaluation offers the ability to use actual events, manufactured events, and simulations to provide a level of proof that is commensurate with the level of definition.

\section{Bibliography}
\bibliographystyle{eptcs}
\bibliography{references}

\begin{thebibliography}{1}
\providecommand{\bibitemdeclare}[2]{}
\providecommand{\surnamestart}{}
\providecommand{\surnameend}{}
\providecommand{\urlprefix}{Available at }
\providecommand{\url}[1]{\texttt{#1}}
\providecommand{\href}[2]{\texttt{#2}}
\providecommand{\urlalt}[2]{\href{#1}{#2}}
\providecommand{\doi}[1]{doi:\urlalt{http://dx.doi.org/#1}{#1}}
\providecommand{\bibinfo}[2]{#2}

\bibitemdeclare{techreport}{DoDSoSguide2008}
\bibitem{DoDSoSguide2008}
\bibinfo{author}{Department \surnamestart of~Defence\surnameend}
  (\bibinfo{year}{2008}): \emph{\bibinfo{title}{Systems Engineering Guide for
  Systems of Systems}}.
\newblock \bibinfo{type}{Technical Report}, \bibinfo{institution}{Department of
  Defence}.
\newblock \urlprefix\url{http://www.acq.osd.mil/se/docs/SE-Guide-for-SoS.pdf}.

\bibitemdeclare{article}{jamshidi2005system}
\bibitem{jamshidi2005system}
\bibinfo{author}{Mo~\surnamestart Jamshidi\surnameend} (\bibinfo{year}{2005}):
  \emph{\bibinfo{title}{System-of-systems engineering-A definition}}.
\newblock {\sl \bibinfo{journal}{IEEE Transactions on Systems, Man, and
  Cybernetics (October 2005)}}.

\bibitemdeclare{article}{maier1998architecting}
\bibitem{maier1998architecting}
\bibinfo{author}{Mark~W. \surnamestart Maier\surnameend}
  (\bibinfo{year}{1998}): \emph{\bibinfo{title}{Architecting principles for
  systems-of-systems}}.
\newblock {\sl \bibinfo{journal}{Systems Engineering}}
  \bibinfo{volume}{1}(\bibinfo{number}{4}), pp. \bibinfo{pages}{267--284},
  \doi{10.1002/(SICI)1520-6858(1998)1:4<267::AID-SYS3>3.0.CO;2-D}.

\bibitemdeclare{article}{manthorpe1996}
\bibitem{manthorpe1996}
\bibinfo{author}{W.H. \surnamestart Manthorpe~Jr.\surnameend}
  (\bibinfo{year}{1996}): \emph{\bibinfo{title}{The Emerging Joint
  System-of-Systems: A Systems Engineering Challenge and Opportunity for APL}}.
\newblock {\sl \bibinfo{journal}{Johns Hopkins APL Technical Digest}}
  \bibinfo{volume}{17}(\bibinfo{number}{3}), pp. \bibinfo{pages}{305--310}.
\newblock \urlprefix\url{http://techdigest.jhuapl.edu/TD/td1703/manthorp.pdf}.

\end{thebibliography}
\end{document}